# CHARGED GRAINS IN SATURN'S F-RING: INTERACTION WITH SATURN'S MAGNETIC FIELD

L. S. Matthews and T. W. Hyde

*Center for Astrophysics, Space Physics, and Engineering Research*
*Baylor University, P.O. Box 97310, Waco, Texas 76798-7310, USA*

## ABSTRACT

Saturn's dynamic F-Ring still presents a challenge for understanding and explaining the kinematic processes that lead to the changing structure visible in our observations of this ring. This study examines the effect of Saturn's magnetic field on the dynamics of micron-sized grains that may become electrically charged due to interaction with plasma in Saturn's rigidly corotating magnetosphere. The numerical model calculates the dynamics of charged dust grains and includes forces due to Saturn's gravitational field, the plasma polarization electric field, a third order harmonic expansion of Saturn's magnetic field, and the F Ring's Shepherding moons, Prometheus and Pandora.

## BACKGROUND

Nearly a quarter of a century ago, man obtained his first clear pictures of one of the jewels of the solar system. Voyagers 1 and 2 sent back dramatic high-resolution images of Saturn's moons and ring system revealing surprises such as the density waves and spokes in Saturn's B Ring. One of the most intriguing puzzles, though, was Saturn's dynamic and ever changing F Ring, first discovered by Pioneer 11 (Gehrels et al., 1980). While Voyager 1 images showed three distinct ringlets with the outer two displaying a kinked and braided appearance (Smith et al., 1981), Voyager 2 images showed a much more regular structure with four separate, non-intersecting strands (Smith et al., 1982). Both Voyager 1 and Voyager 2 detected brighter clumps within the rings with a temporal dependence ranging from days to months (Showalter, 1997).

Several studies attempted to explain these unique features through gravitational interactions with the F Ring's shepherding moons or imbedded moonlets, while others looked at the possible interactions between charged grains and Saturn's magnetosphere. Twenty years later, as we await the arrival of Cassini, the latest probe sent to Saturn, the factors governing the dynamics of the F ring are still unclear.

Detailed photometric studies of the data collected by the Voyager probes indicated that the F Ring consisted of a 1 km core of centimeter-sized particles, with an envelope of micron and sub-micron material extending inward approximately 50 km (Showalter et al., 1992). The Hubble Space Telescope was able to glean additional information from a stellar occultation during Saturn's ring plane crossing in 1995. The data from this study corroborated the evidence for the existence of a core with an interior envelope, but suggested that the majority of the material was 10 μm or larger with a lower cutoff in size of 0.3-0.5 μm. Additionally, an upper limit for the amount of submicron material was set at 28% (Bosh et al., 2002). It has since been postulated that the difference in the two data sets may be due to temporal variations in the ring structure.

The F Ring lies within the inner part of Saturn's magnetosphere, which contains plasma that rigidly corotates with the planet (Grün et al., 1984), and thus the dust in Saturn's F Ring can become charged. Since the plasma parameters in the vicinity of the F Ring are poorly constrained, the magnitude of the charge on the dust grains is not well known. Estimates of the charge on micron and submicron grains within the F Ring have ranged from less than one electron (Showalter and Burns, 1982) to q = C(–38 V) (Mendis et al., 1982), where $C \approx 4\pi\varepsilon_o a e^{-a/\lambda_D}$ is the capacitance of a spherical grain of radius *a* immersed in a plasma with Debye length $\lambda_D$ (Whipple et al., 1985).

Charged grains' orbits will be perturbed by the planetary magnetic field with the magnitude of this perturbation depending primarily on the grain's charge-to-mass ratio. In addition, the gravitational interaction of a nearby satellite with a narrow ring produces a wave downstream of the moon. Eccentricities in the orbit of either the satellite or the ring can produce azimuthal clumping, having a spatial periodicity initially equal to the wavelength due to the perturbing satellite (Kolvoord et al., 1990). The purpose of this study is to examine the interaction of weakly charged grains with Saturn's magnetic field and the F Ring's shepherding moons, Prometheus and Pandora. Complications not addressed in this study are the inclination of the F Ring and satellite orbits, collisional damping between ring particles, collective effects of the charged grains and plasma, and the effect of moonlets imbedded within the ring.

**NUMERICAL MODEL**

**Box_Tree Code**

The numerical code employed in this study, box_tree, has been used to model a variety of systems consisting of both charged and uncharged grains in ring systems (Richardson, 1993, 1994), protoplanetary clouds (Richardson, 1995, Swint et al., 2002), and laboratory dusty plasmas (Vasut and Hyde, 2001, Qiao and Hyde, 2003). It is a combination of a box code, which specifies the boundary conditions and allows the equations of motion to be linearized, and a tree code, which allows a rapid calculation of interparticle forces. The code has been modified to include a spherical harmonic expansion model of Saturn's magnetic field as well as the gravitational potentials of the F Ring's shepherding moons, Prometheus and Pandora.

Saturn's Magnetic Field

The magnetic field around a planet can be regarded as the sum of the contributions from the planetary dynamo (internal sources) and the exterior plasma sheet produced by the interaction of the solar wind with the planetary magnetic field (Connerney, 1993). Generally, for points to the interior of the plasma sheet, this contribution is negligible when compared to the field due to internal sources and may be disregarded.

The magnetic field can thus be calculated at any given point using a spherical harmonic expansion model where the magnetic field is defined as the gradient of a scalar potential. The coefficients in the spherical harmonic expansion are determined by best fits to experimental data from *in situ* measurements made by Pioneer 11 and Voyagers 1 and 2. A number of researchers have presented various best-fit models; the most accurate of these for Saturn's magnetic field is the $Z_3$ model (zonal harmonic of degree 3), an axisymmetric octupole with the rotational and magnetic axes aligned to within 0.1º (Connerney, 1993). This is the model employed by this study.

Shepherding Satellites

The effect of the shepherding satellite is included as an external gravitational potential in the rotating frame. The center of the box containing the ring particles is on a circular orbit, by construction. Thus, the combined eccentricities of the ring and satellites (Bosh et al., 2002; McGhee et al., 2001) are mapped back to the satellites alone, following the procedure of Showalter and Burns (1982).

An "eccentric displacement vector" **E** is defined with polar coordinates ($ae$, ω) where *a* is the maximum radial displacement out-of-round for a given body, *e* is the eccentricity, and ω is the longitude of the pericenter. Any ring/satellite configuration can thus be described by $a_0$, $a_1$, $\mathbf{E_0}$, and $\mathbf{E_1}$, where the subscripts zero and one refer to the ring and satellite, respectively. The effect of the satellite on the ring depends on $\mathbf{E_1} - \mathbf{E_0}$. By choosing to set $\mathbf{E_0}' = 0$ and $\mathbf{E_1}' = \mathbf{E_1} - \mathbf{E_0}$, all of the eccentricity can be assigned to the satellite. The effect of the "primed" satellite on a circular ring is essentially the same as that of the unprimed scenario, in which both the satellite and the ring have eccentric orbits. The reference longitude for the satellite's eccentricity is set at run time, so that the position of the satellite can be varied from closest approach to furthest approach as it passes the center of the box.

Boundary Conditions

The box_tree code makes use of periodic boundary conditions. In the general case, ghost boxes surround the central box and particles that are near a boundary interact with ghost particles in the neighboring box. As a particle leaves the central box, it is replaced by its ghost particle entering from the opposite side, keeping the total number of particles in the simulation constant (Richardson, 1993). These boundary conditions were modified slightly to take advantage of special properties of the narrow F Ring. The box size was chosen to be much greater than the radial width and vertical thickness of the ring. Thus particles are only able to leave the box in the azimuthal direction, requiring only two ghost boxes (Figure 1). Because the box is a non-inertial, rotating frame, particles that are displaced inward toward Saturn (-x direction) have a positive azimuthal drift (in the +y direction) while particles displaced outward (+x direction) have a negative azimuthal drift.

Since Prometheus, the inner shepherding satellite, leads the box, particles that are displaced inward and travel faster than the mean motion of the box will spend more time in its vicinity. Likewise slower particles will spend more time in the vicinity of Pandora, the outer satellite. This effect was modeled by tracking the number of azimuthal boundary crossings, *n*, for each particle. The number of boundary crossings times the length of the box, *nL*, is added to the azimuthal position of the grain, and this adjusted position is used in calculating the grain-satellite separation and the gravitational force due to the satellite. The grain's unadjusted position within the box is used in calculating the effect of the azimuthally symmetric magnetic field.

**Equations of Motion**

The acceleration of a charged grain in a planetary magnetic field, **B**, and gravitational field is

$$\ddot{\mathbf{R}} = \frac{q}{m}(\mathbf{E} + \dot{\mathbf{R}} \times \mathbf{B}) - \frac{GM_p}{R^3}\mathbf{R} - \nabla\phi \qquad (1)$$

where $M_p$ is the mass of the planet, **R** is the distance from the center of the planet to the grain, and *q* and *m* are the grain's charge in coulombs and mass in kilograms, respectively. The grain-grain gravitational and electrostatic interactions are included in the $-\nabla\phi$ term, and are calculated by the tree code. The electric field, $\mathbf{E} = -(\mathbf{\Omega}_p \times \mathbf{R}) \times \mathbf{B}(\mathbf{R})$, results from the relative motion of the ring particles with respect to the magnetic field corotating with the angular velocity of the planet, $\mathbf{\Omega}_p$.

The equations of motion can be linearized in the box frame, which is rotating about the planet with constant angular velocity $\mathbf{\Omega}_k = \Omega_k \hat{z}$ with magnitude $\Omega_k = \operatorname{sqrt}(GM/R_b^3)$. $R_b$ is the distance from the center of the box to the planet center and M is the mass of the central planet. Using the fact that $R_b \gg r$, the position of the particle within the box, the equations of motion can be linearized by expanding about $R = R_b(1 + r/R_b)$ and are given by

$$\ddot{x} = F_x + 3\Omega_k^2 x + 2\Omega_k^2 \dot{y} + \ddot{x}_m, \qquad \ddot{y} = F_y - 2\Omega_k^2 \dot{x} + \ddot{y}_m, \qquad \ddot{z} = F_z - \Omega_k^2 z + \ddot{z}_m \qquad (2)$$

where $\mathbf{F} = -\nabla\phi$ is the sum of the gravitational and electrostatic forces per unit mass due to all other particles and the accelerations due to the magnetic field in each dimension are given by

$$\ddot{x}_m = \frac{q}{m}\left[(\Omega_k - \Omega_p)R_x B_z + \dot{y}B_z - \dot{z}B_y\right], \qquad \ddot{y}_m = \frac{q}{m}\left[\dot{z}B_x - \dot{x}B_z\right],$$

$$\ddot{z}_m = \frac{q}{m}\left[-(\Omega_k - \Omega_p)R_x B_x + \dot{x}B_y - \dot{y}B_x\right]. \qquad (3)$$

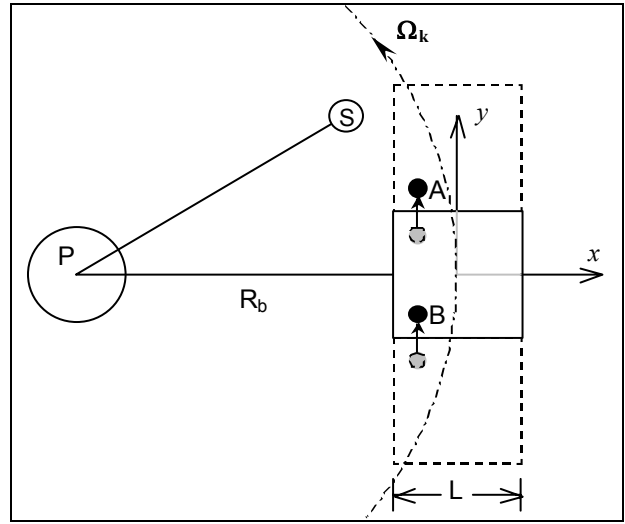

Fig. 1. Boundary conditions in the rotating coordinate system. As particle A crosses the boundary of the central box, it is replaced by its ghost particle, B. The position at B is used to calculate the force due to the azimuthally symmetric magnetic field. The position at A, or the number of boundary crossings *n* times the box length *L*, is used in determining the distance between the particle and the shepherding satellite, S.

**NUMERICAL SIMULATIONS**

**Epicyclic Motion**

The theory of the motion of charged dust grains within a rigidly corotating magnetosphere was developed by Mendis et al. (1982) and termed "gravitoelectrodynamics" (GED). The motion of the charged grain due to interactions with the magnetosphere is treated as a perturbation to the Keplerian orbit. Highly symmetric dipole fields, such as Saturn's, produce a perturbation which may be described as an elliptical gyration about a guiding center moving on a circular orbit with angular velocity $\Omega_G$. The ratio of the minor to major axis of the ellipse depends on the charge-to-mass ratio of the grains, approaching ½ as q/m → 0 and unity as q/m increases. The radial components of the magnetic field also produce oscillations out of the plane (Xu and Houpis, 1985).

The individual orbits of weakly charged grains with varying charge-to-mass ratios were tracked to determine the magnitude of the perturbation caused by Saturn's magnetic field. Each grain was given an initial zero velocity at the center of a 200 km box, centered at the mean orbital distance of the F Ring, and orbiting the planet at the mean Keplerian speed of the F Ring, $\Omega_k = 1.168 \times 10^{-4}$ rad/s. The subsequent motion of the grains was tracked for several orbital periods without the gravitational interactions of the shepherding moons.

The radial excursion of a charged grain from its initial position within the ring scales directly with its charge-to-mass ratio. The resulting orbits, adjusted for the motion of the guiding center, are shown in Figure 2 for grains with charge-to-mass ratios ± 0.01 C/kg to ± 0.04 C/kg. As can be seen, grains with |q/m| = 0.01 C/kg have a maximum radial excursion of about 10 km, while that for grains with |q/m| = 0.04 C/kg is greater than 60 km. Both the Voyager and HST data indicate that the radial width of the F Ring envelope is 50 km (Showalter et al., 1992; Bosh et al., 2002). Requiring that charged grains be confined to the envelope thus sets a maximum |q/m| = 0.03 C/kg for grains in this region. Another striking feature of the envelope is that it is displaced inward from the F Ring core. If the postulate is true that submicron and micron-sized grains produced in the core are the source of the material in the F Ring envelope, these grains must be negatively charged, as only negatively charged grains are swept inward by Saturn's magnetic field. The magnitude of the out-of-plane oscillation is on the order of 3 km for a grain with q/m = ±0.04 (Figure 3). This is in good agreement with the ring thickness range of 2-5 km derived from stellar occultation data (Bosh et al., 2002).

**Effects of Shepherding Moons**

The gravitational perturbations of the shepherding moons were investigated for a population of 5000 grains within a 2800 km box. This is equivalent to modeling an entire ring with $1.5 \times 10^6$ grains. The grain radii ranged from 0.5 μm to 10.0 μm and followed a power law distribution with q = 3.5. Grains were initially established on stable orbits within the 50 km F Ring envelope without perturbations from shepherding moons. The charged grains had epicyclic orbits about a guiding center while uncharged grains had local Keplerian orbits. Estimates of plasma temperature and density in the F Ring range from 10 eV < $kT_e$ < 100 eV (Grün et al., 1984) and 10 cm$^{-3}$ < $n_o$ < 100 cm$^{-3}$ (Mendis and Rosenberg, 1996), respectively. For this study, the plasma temperature and density were assumed to be constant across the width of the box. Using the values $kT_e$ = 20 eV, $n_o$ = 100 cm$^{-3}$, and dust density $N_d$ = 30 cm$^{-3}$ (Mendis and Rosenberg, 1996), calculated grain charges yield q/m = -3.8 × 10$^{-6}$ C/kg for $a_{max}$ and

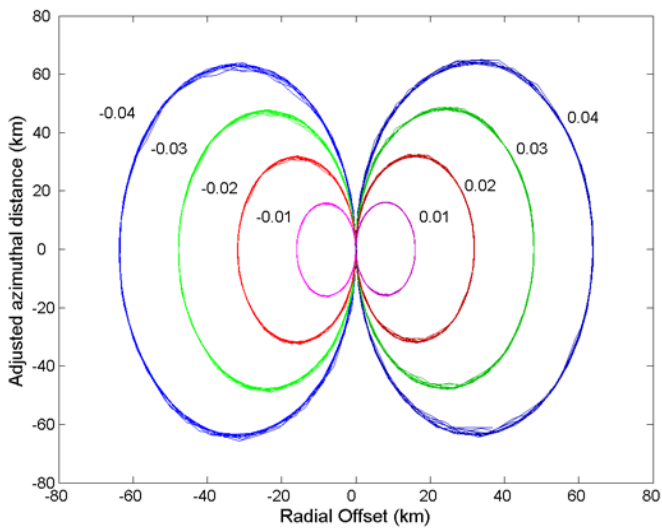
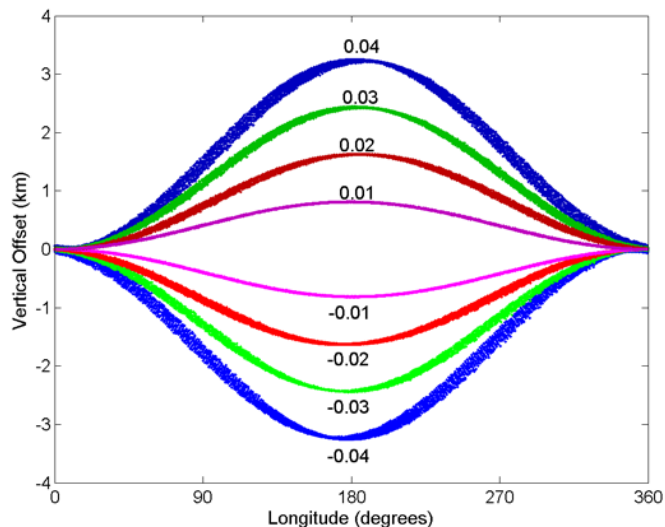

Fig. 2. Epicyclic orbits of charged grains over 10 orbital periods. The x-axis is the radial offset relative to the center of the F Ring. Azimuthal positions are adjusted for the mean motion of the guiding center. The size of an epicycle depends on a particle's charge-to-mass ratio, as indicated by the labels.

Fig. 3. Vertical motion of charged grains during 100 orbital periods about Saturn. Grains initially centered in the F Ring core undergo simple harmonic motion in the direction normal to the equatorial plane due to the northward offset of Saturn's dipolar field. Labels indicate the charge-to-mass ratio of a grain.

$q/m = -0.03$ C/kg for $a_{min}$ (Barge, 2002). It should be noted that grain charges vary only slightly for $kT_e < 25$ eV and as a result the relative dust to plasma density ratio has a relatively small effect on the qualitative results of this study (Barge, 2002).

Ring interactions were modeled with Prometheus and Pandora passing the box both singly and simultaneously. The data from multiple consecutive boxes was concatenated to form a larger ring section. The adjusted azimuthal positions were used in plotting the ring configuration with overlapping data from individual boxes.

The ring configurations before and after passage of Prometheus are shown in Figure 4 for both charged and uncharged grains. In all three cases with uncharged grains, the passing moon(s) excited waves in the ring as shown in previous models (e.g. Showalter et al. (1992)) with the grains remaining in phase across the entire size range. However a distinct phase difference evolves for charged grains with the larger charged grains ($a \geq 3.0$ μm) having very similar orbits to uncharged grains. The smaller charged grains ($a < 1.0$ μm) were more strongly affected by the inner satellite than the larger grains. As a result, a clear difference can be seen between the behavior of the charged and uncharged grains.

The resulting grain positions for an extended run with both Prometheus and Pandora is shown in Figure 5. During the simulation, each satellite made three passes of the ring section, with the two moons first passing the ring section during the time interval of $1.0 \times 10^5$ to $1.7 \times 10^5$ s. The larger grains with low charge-to-mass ratios clearly show periodic clumping with spacing on the order of 10,000 km, which is analogous to structure observed in the Voyager data (Kolvoord et al., 1990). The smaller grains, while exhibiting the waves produced by interaction with the shepherding satellites, do not form distinct clumps.

**CONCLUSIONS**

It has been shown that even weakly charged grains significantly affect the dynamics of the F Ring. Sub-micron grains that obtain a small (negative) charge are swept inward from the core by Saturn's magnetic field. A maximum charge-to-mass ratio $|q/m| = 0.03$ C/kg has been established for grains with stable orbits within the F Ring envelope. This is easily attainable for even low plasma densities in the F Ring; for example, a charge of one electron on a 0.1μm ice grain will yield this value. Gravitational interactions with the shepherding moons are also affected by the presence of charged grains. The different size populations exhibit phase differences in the wavy orbits imposed by the shepherding moons, with these phase differences perhaps playing a role in the formation of the braids observed in the ring. Further separation based on size is also observed in the formation of clumps, which consist mainly of the larger grains while the smaller grains are dispersed evenly about the ring. While this work is preliminary, it does point to possible interesting macroscopic effects. The difference in size populations mentioned above may be detectable in the Voyager optical depth data taken at different wavelengths and should also be seen in the data gathered by several of the instruments on the Cassini mission which will reach Saturn in 2004.

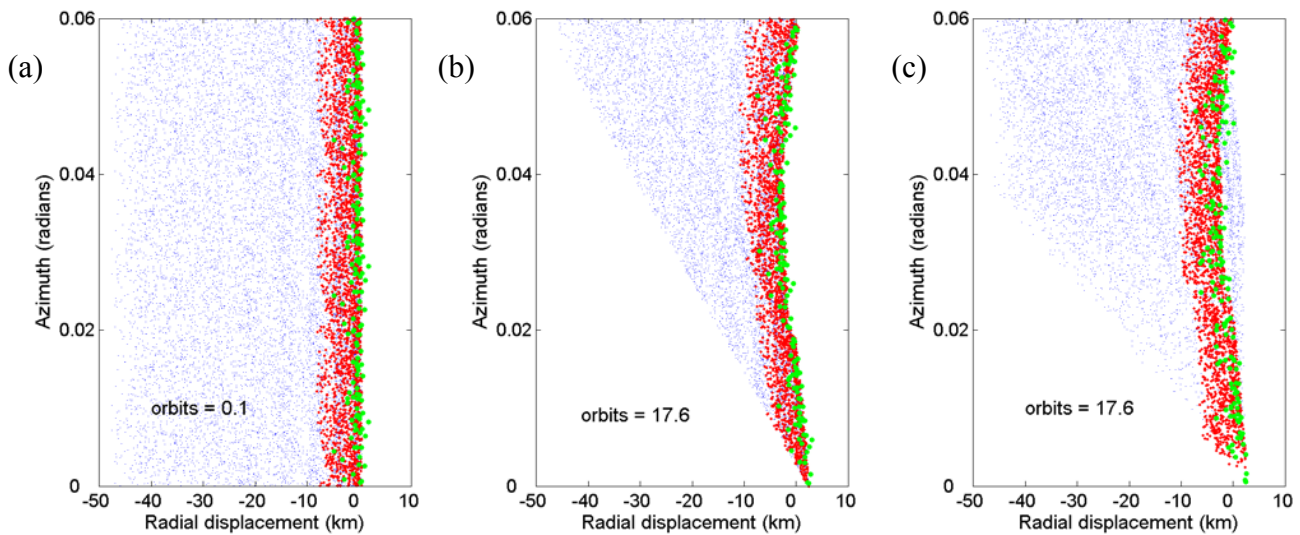

Fig. 4. Ring configuration for interaction with Prometheus. Grains with $r < 1.0$ μm are blue, $1.0$ μm $\leq a < 3.0$ μm are red, and $a \geq 3.0$ μm are green. The initial configuration of the ring is shown in (a). Uncharged grains remain in phase (b), while a phase difference is visible for charged grains between the different size populations (c).

Future work in this area will include additions to the code, such as angular momentum and energy transfer between the shepherding moons and ring. A parallel version of the code is under development that will allow data for higher dust densities to be compared with optical depth data.

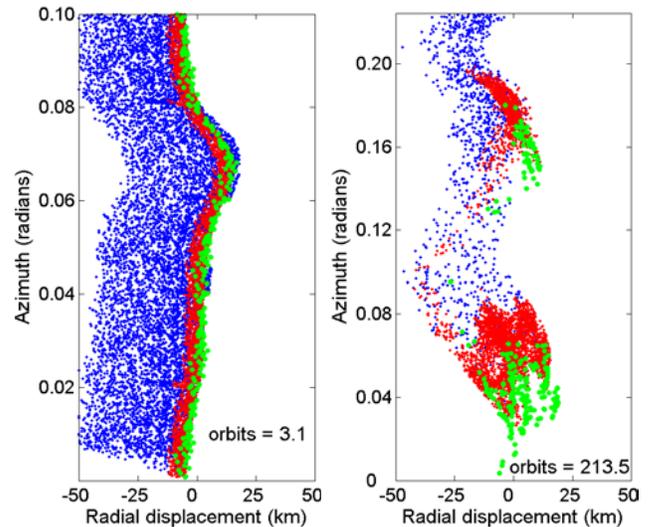

Fig. 5. Ring configuration (a) after Prometheus and Pandora's first pass and (b) after completion of three passes by each moon. The number of orbits refers to the motion of the box about Saturn.

E-mail addresses: Lorin_Matthews@baylor.edu, Truell_Hyde@baylor.edu